\documentclass{article}
\usepackage[margin=1in]{geometry}
\usepackage{graphicx}
\usepackage[toc]{appendix}
\usepackage{amsmath} 
\usepackage{amsfonts}
\usepackage{amsthm}
\usepackage[lofdepth,lotdepth]{subfig}
\usepackage{enumitem}
\usepackage{multirow}
\usepackage{url}

\newtheorem{thm}{Theorem}[section]
\newtheorem{lem}[thm]{Lemma}

\newtheorem{cor}{Corollary}

\setlist[itemize]{leftmargin=*}

\providecommand{\keywords}[1]
{
  \small	
  \textbf{\textit{Keywords---}} #1
}

\title{Reputation-based PoS for the Restriction of Illicit Activities on Blockchain: Algorand Usecase }

\author{Mayank Pandey, Rachit Agarwal,
Sandeep Kumar Shukla, Nishchal Kumar Verma\\
		IIT Kanpur,	Kanpur, India \\
	\texttt{\{pandeym,rachitag,sandeeps,nishchal\}@iitk.ac.in}} 
\begin{document}
\maketitle

\begin{abstract}
In cryptocurrency-based permissionless blockchain networks, the decentralized structure enables any user to join and operate across different regions. The criminal entities exploit it by using cryptocurrency transactions on the blockchain to facilitate activities such as money laundering, gambling, and ransomware attacks. In recent times, different machine learning-based techniques can detect such criminal elements based on blockchain transaction data. However, there is no provision within the blockchain to deal with such elements. We propose a reputation-based methodology for response to the users detected carrying out the aforementioned illicit activities. We select Algorand blockchain to implement our methodology by incorporating it within the consensus protocol. The theoretical results obtained prove the restriction and exclusion of criminal elements through block proposal rejection and attenuation of the voting power as a validator for such entities. Further, we analyze the efficacy of our method and show that it puts no additional strain on the communication resources.   
\end{abstract}


% \section*{Note}
% \textbf{This work has been submitted to the IEEE for possible publication. Copyright may be transferred without notice, after which this version may no longer be accessible.}

\section{Introduction}~\label{sec:intro}

Digital payments saw an exponential rise in 2020 due to COVID-19 and work-from-home culture. This also saw a rise in cryptocurrency-based transactions with market capitalization tripling in 2021~\cite{IMFgfsr}. Such a rise in adoption and demand not only saw the prices of various cryptocurrencies rise but also saw an increase in illicit activities~\cite{ChainAnalysis2021}.

Illicit activities are performed mainly due to the exploitation of the vulnerabilities in the blockchain infrastructure (hardware/software)~\cite{tanmayVulnerability2021}. For example, in the case of Ethereum, the DAO hack exploited Reentrancy vulnerability~\cite{SamReentrancy} where DAO lost $\$70$ Million. In a recent Poly network attack, the attacker exploited the multi-sig vulnerability to siphon-off $\$600$ Million \cite{PollyAttack}. Apart from the aforementioned types of attacks, attackers also target cryptocurrency users through social engineering techniques. Additionally, some of the users use cryptocurrency to facilitate different illicit activities such as gambling and money laundering. In terms of impact, there was a $311\%$ rise in ransomware activity (a social engineering-based attack) that caused $\$20$ Billion loss alone in 2020~\cite{ChainAnalysis2021}. %According to various reports, 
Activities such as terrorist financing, darknet marketplaces, and financing of child abuse material are also prominent in cryptocurrency-based blockchains.

Thus, a question we ask is \textit{how can we restrict such illicit activities in cryptocurrency-based blockchain?} There are many machine learning (ML) based techniques that have been proposed to detect suspicious accounts are not only based on the transaction history (behavior based)~\cite{Agarwal2021, masterstheBan, agarwal2020Adver, masterstheDep} but also utilise the meta-data information attached to the address~\cite{sachan2021identifying}. Further, services such as Chainalysis\footnote{\url{https://www.chainalysis.com/}} perform a dynamic exploration of the blockchain through use of proprietary ML algorithms. Such services and ML-based approaches are not integrated with the blockchain infrastructure. They operate outside the blockchain and have characteristics such as using the transaction history, being paid, dependent on ground-truth information, and are computationally expensive. Moreover, few approaches define reputation of user in the blockchain network using various mechanisms that although are  behavior oriented but are either communication expensive~\cite{9045608}, maintain a side chain~\cite{aluko2021proof}, or biased~\cite{9355472}.

Within the blockchain network, another set of approaches exist that design and modify consensus algorithm to limit illicit activities. For example, in Ethereum~\cite{ethslash}, \textit{slashing} is proposed to confiscate validating user's stake in case the user is involved in any malicious activity (such as signing two proposed blocks at the same time and validating double-spending transactions). However, the definition of illicit activities for these algorithms is limited to the activities that attempt to disrupt the functioning of the blockchain network by exploiting vulnerabilities in the blockchain system. For illicit activities such as money laundering and other social engineering based attacks such as phishing, there are no countermeasures except for Anti-Money Laundering (AML) regulations outside the scope of blockchain. There is no inbuilt mechanism in the blockchain to limit criminals from exploiting cryptocurrency-based blockchains for their illicit gains. Even if a certain user practices/shows illicit behavior, it takes a lot of time for the Law Enforcing Agencies (LEAs) to capture him~\cite{IBMreport}, sometimes which never happens due to pseudonymity in the blockchain networks. Thus, justifying the need for regulation in the cryptocurrency-based blockchains for the users involved in illicit activities. The gap present defines our objective as to propose a mechanism for regulating malicious entities within the framework of blockchain. In this work, we provide a mechanism to deal with the user accounts involved in illicit activities in the consensus process and conduct a theoretical feasibility study for the same.

For our purpose, we select the Proof of Stake (PoS) consensus algorithm as implemented in Algorand~\cite{chen2019algorand}. Our choice is based on the following facts. It provides the deterministic finality of the block immediately after the end of each block consensus round as its PoS algorithm combines with the byzantine fault tolerance (BFT). The creators of Algorand claim to solve the blockchain trilemma~\cite{blocktrilemma} of decentralization, security, and scalability. The features pertaining to the proposer/validator selection process and data encryption claim the Algorand blockchain to be most resilient against the adversarial attacks on the network among the existing permissionless blockchains. Even though Algorand claims limiting illicit activities, it has no inbuilt provision against the users engaged in social engineering-based illicit activities. We thus propose modifications to Algorand's PoS consensus algorithm and perform a theoretical analysis of our method.

Our method integrates the result of a state-of-the-art ML model (that identifies the probability of being suspicious, equivalent to the 1 - \textit{reputation score})~\cite{Agarwal2021} with Algorand's consensus algorithm. In Algorand, to achieve consensus, each node performs several steps (cf. Appendix~\ref{app:algo}). In the first step, a node depending on certain criteria, forms and proposes a block. This block is then validated by other nodes (called validators, selected based on certain criteria) in multiple steps powered by a verifiable random function (VRF). We parameterize the received output of the VRF with the reputation score to make the decision on accepting or rejecting a block proposal. The votes of the validators are changed in accordance with their respective reputation score. This process happens at each node with no additional communication of any extra information. Each user makes a decision based on the reputation scores unilaterally and then the cumulative effect of all such decisions leads to the outcome of consensus round. Our methodology restricts the criminal entities and, in the process, freezes their accounts in a way that no transactions from such entities will get approved by an honest validator. We explain our proposed method in detail in section~\ref{sec:algo}.
 
Note that our method is applicable to other cryptocurrency-based blockchains as well but with suitable modifications according to the consensus process used. Our approach is different from slashing. In our method, there is no loss of cryptocurrency. Instead, there is a temporary loss of voting power based on the behavior. Also, note that we do not focus on detecting the vulnerabilities in the system but instead focus on restricting illicit activities that are either socially motivated, exploit vulnerabilities, or use cryptocurrency as a tool to facilitate the illicit activities.

In summary, our contributions are:
\begin{itemize}
    \item \textit{\textbf{State-of-the-Art Analysis}}: We identify the gaps within the different consensus algorithms with respect to the handling of users carrying out illicit activities.
    \item \textit{\textbf{Reputation score based Consensus method}}: We propose an algorithm for associating reputation to each user and use it to enhance resilience against criminal activities in PoS consensus-based blockchain networks.
    \item \textit{\textbf{Feasibility Analysis}}: We incorporate our method with the Algorand blockchain network consensus process and theoretically analyze the same 
    in terms of parameters such as communication hop, time, space and overhead. 
\end{itemize}

The remaining part of the manuscript is organized as follows. Section~\ref{sec:background} provides a brief overview of the background of relevant blockchain functionalities and threats to the network. Section~\ref{sec:RW} lists out the related work with respect to detection of illicit users and proposed reputation models followed by the motivation for our proposed methodology. In section~\ref{sec:algo}, we provide the proposed methodology for Algorand blockchain in detail along with the derived theoretical results. In section~\ref{sec:analysis}, we analyze the proposed methodology with respect to communication and space parameters to show the effect on the Algorand consensus protocol functioning. Finally, in section~\ref{sec:discussion}, we conclude along with a discussion into the future prospects for our proposed approach.

\section{Background}\label{sec:background}

In this section, we focus on the consensus process in permissionless blockchain networks. In a permissionless blockchain network, every node has the opportunity to perform a transaction, propose a block of transactions or verify and validate the same. The communication in all permissionless blockchains' consensus protocols is the variant of the same in Byzantine Fault Tolerance (BFT) consensus process~\cite{10.1007/978-3-540-39878-3_19}. However, for cryptocurrency-based permissionless blockchains, the functioning of the BFT protocol is modified. There are several consensus protocols such as Proof of Work (PoW), and Proof of Stake (PoS), which are examples of such variations. Here, we only discuss the major consensus protocols that are adopted especially by the cryptocurrency-based blockchain and provide an overview of different threats and illicit activities that have happened in such cryptocurrency-based blockchain. For the detailed study of the same, in~\cite{Bano2019}, the authors survey different consensus protocols used in different blockchains. They highlight the performance and scalability issues in different protocols and emphasize the need to mitigate these issues for enabling widespread adoption. For the scope of our paper, we provide a brief overview.  

\subsection{Different Consensus Mechanisms} 

Blockchain technology uses consensus mechanism for reaching the decision over the induction of new blocks into the chain. There are five stages of engagement between the blockchain network users during the consensus process~\cite{8972381}. These are proposing block, information transmission, validating the block components, finalizing the block, and providing the incentive. The primary objective of a consensus procedure is to accept or reject the transactions happening between the users of the blockchain network. The decision of selection/rejection must get agreed upon by the majority of validating users present in the network. It is only after the majority validation, the block with the specific transactions gets added to the chain. The majority condition is defined to take into account the presence of malicious nodes. The blockchain networks operate on the assumption that the majority of nodes are honest ($51\%$ for PoW, $\frac{2}{3}$ for BFT based consensus).

\subsubsection{BFT Consensus}

BFT consensus~\cite{10.1007/978-3-540-39878-3_19} is referred to as the consensus protocol, which is resilient against network crash in the presence of faulty and malicious nodes. For a BFT consensus protocol to work, at least $\frac{2}{3}$ of the total number of nodes in the given network must be honest, non-faulty, and agree on the same output to reach consensus. The BFT consensus provides output with deterministic finality. BFT by structure demands multiple rounds of message communication between all the nodes to reach a consensus. Due to the extensive communication requirement, a decentralized network using BFT based consensus is not scalable. Some permissionless blockchain networks also implement Hybrid versions of BFT, such as PoW-BFT for Byzcoin \footnote{\url{https://actu.epfl.ch/news/byzcoin-an-innovative-solution/}} and BFT based PoS for Algorand. 

\subsubsection{Proof of Work (PoW)}

In PoW, a miner, to propose a block of transactions solves a mathematical problem to identify a suitable random value known as the nonce. Once a user identifies the nonce for its proposed block, it transmits the proposed block information to every other node in the network for verification and subsequently validation. When the proposed block gets validated by the majority of the network participants, it gets added to the blockchain. The creator of the proposed block is credited with a reward. The only known way to do mining is guessing the suitable nonce value through the brute force method. It provides a fair chance in proportion with computing resources for every participating node to add the new block into blockchain successfully.

Since its conception, the Bitcoin blockchain network has expanded considerably in terms of both network participants and transactions volume. However, it has some drawbacks, such as the high requirement of computational power and energy resources for mining. Also, except for the work of participants successful in forming the block, the rest of the mining work is complete wastage as it does not generates any useful output or utility. At present, Bitcoin blockchain network is no longer decentralized due to the domination of mining pools and specialized mining equipment. All the factors mentioned above prompted the research to find resource and energy-efficient consensus protocols.  

\subsubsection{Proof of Stake (PoS)}

In PoS consensus protocol, the probability of getting credited with the block formation is proportional to the amount of cryptocurrency staked for the consensus procedure. PoS is resource-efficient in comparison to PoW due to low computational requirements. The biggest proponent of PoS is the Ethereum blockchain, which is transitioning from its original PoW based consensus. However, PoS has shortcomings specific to the stake-based methodology. The most significant limitation is its favor to the highest stakeholders. The users with high stakes in the network can propose multiple blocks, also referred to as the ``nothing-at-stake'' problem. 

There are different variants of PoS, such as hybrid PoW/PoS, Delegated Proof of Stake (DPoS), and committee-based PoS. Table~\ref{tab1} lists the advantages and limitations associated with such methods. The hybrid version of PoS/PoW works by adjusting a user's PoW threshold using its cryptocurrency stake in the network. In DPoS, the nodes not having enough amount of cryptocurrency can participate in the validation process by taking part in the selection of committees amongst the aspiring committee groups. In committee-based PoS, the block validating committee for upcoming consensus rounds is selected using a pseudo-random number generator. The output of the generator depends on the stakes of the nodes. Each selected committee has a leader who proposes the block. The proposed block has to be validated by the majority of remaining members of the same committee. However, in committee-based PoS, a malicious node can carry out denial of service (DoS) attacks on the committee members and leaders to disrupt the network. Therefore, it is imperative that the honest nodes possess more than $50\%$ of cryptocurrency ($66\%$ in case of BFT-PoS consensus methods). 

In the aforementioned consensus methods, the block finality is probabilistic. The transactions in a block get confirmed after a sufficient number of new blocks get added after it into the chain. A combination of PoS with byzantine fault-tolerant (BFT) consensus is used like in Algorand~\cite{chen2019algorand} to make block finality deterministic, i.e., transactions finalized as soon as the block gets added. In BFT based PoS, block proposing is done based on the users' stake while the rest of the procedure is followed through BFT based communication.  

\subsubsection{Other Consensus Methodologies}

There are other consensus algorithms that have been used in several other permissionless blockchains to remove the disadvantages posed by PoW and PoS.  For example, Proof of Retrievability (PoR)~\cite{juelsretrievability} consensus algorithm. It is also referred to as Proof of Space. Instead of computational resources requirements like PoW, here, a user's chances of block formation are in proportion with the amount of local space available committed to the network as stake. It is used in Permacoin cryptocurrency blockchain~\cite{permacoin}. A consensus protocol called Ripple~\cite{schwartz2014ripple} is used in XRP cryptocurrency. However, in Ripple, the validating nodes and client nodes are predefined. Each validator has a list of trusted validators known as unique node list (UNL). The validating votes are collected for each transaction from the UNL peers before inducting them into the block. Further, at least $80\%$ yes votes are required for the transactions to go through validation. 

Apart from the aforementioned consensus protocols, there is a separate class of consensus protocols for blockchains where the ledger is modeled as a directed acyclic graph (DAG). These blockchains either have blocks as vertices in the DAG or have transactions as vertices in the DAG. In block-based DAG, each block is hash linked to multiple parent blocks instead of just a single one as in the case of traditional blockchains such as Bitcoin and Ethereum. The selection of parent blocks for the same is a significant issue for every new block. Unlike serially linked blockchains, multiple blocks get added at different points, and therefore, the problem of conflicting transactions may arise. There are two consensus protocols proposed based on block-based DAG ledger, PHANTOM~\cite{2018phantom} and SPECTRE~\cite{2016spectre}. Both follow PoW based block proposal and validation. The transaction-based DAG blockchain is adopted through its consensus protocol Tangle~\cite{popov2018tangle}. The tangle protocol is used by IOTA, a cryptocurrency-based distributed ledger designed for IoT devices. In IOTA, a node can add a new transaction after verifying at least two existing transactions and attaching its transaction to these two. The attachment is made by including the hash values of the selected transactions for validation and solving the PoW puzzle to broadcast along with the transaction. Since cryptocurrency is involved, the malicious entities can carry out feature-based exploitation of IOTA. The IOTA tangle provides protection against Denial of service and spam attacks, but not against social engineering based attacks. 

\begin{table}[htbp]
\caption{Permissionless Blockchain Consensus Protocols.}\label{tab1}
    \resizebox{\columnwidth}{!}{\begin{tabular}{|l|l|l|l|l|}
\hline
\multirow{3}{*}{Consensus Protocol} & \multirow{3}{*}{Advantages} & \multirow{3}{*}{Limitations} & \multirow{3}{*}{Blockchain} & Protection against \\
 &  &  &  & social engineering \\
  &  &  &  &  attacks \\
\hline
\hline
PoW  &  Tampering resistant & High computational & Bitcoin & None\\
(Nakamoto) \cite{nakamoto2008bitcoin}& No double spending &  resource requirement & &\\
 & $50\%$ fault tolerance & High energy demand & &\\
\hline
PoW (GHOST) \cite{buterin2013ethereum}& Orphaned Blocks & High resource & Ethereum & None\\
&included and rewarded & requirement & &\\
\hline
Hybrid PoW-BFT \cite{Byzcoin}& Stabilized consensus & Higher communication & Byzcoin &None\\
& & overhead & &\\
\hline
Chain-based & Less resource wastage & Nothing-at-stake problem & Peercoin &None\\
PoS \cite{KingPeercoin}& & Wealth Centralization & &\\
\hline
Committee-based & Pseudo-random selection, & Scalability for  & Ourosboros, &None\\ 
PoS \cite{OuroborosPraos}& Orderly block generation, &  large committees, & Ouroboro &\\
& Predefined consensus& Targeted attacks &  praos&\\
& time period& on committee members  &  &\\
%&  &  &  &\\
%&  & & & \\
\hline
BFT-based PoS \cite{chen2019algorand} & Deterministic finality & $66\%$ honest majority& Algorand &None\\
 & Fair reward distribution & requirement & Tendermint &\\
\hline
Delegated PoS \cite{kwon2019cosmos}& Block reward option & Collusion of & Cosmos& None\\
 & for non-validators & malicious delegates & &\\
\hline
Proof of  &Less computational & Malicious users & Permacoin& None\\
 Retrievability \cite{juelsretrievability}& wastage & with large storage & &\\
\hline
Ripple consensus & Better performance &$80\%$ voting threshold & XRP &Partial\\
protocol \cite{schwartz2014ripple} & High level of trust & High centralization& &\\
\hline
Tangle \cite{SILVANO2020307} &Individual transactions & Conflicting transactions & IOTA &None\\
 & direct addition, Coordinator & Lot of indirect confirmations& &\\
 & checking transactions & required, Coordinator compromises with & & \\
 & Provision for less storage & decentralization in blockchain trilemma & & \\
\hline
\end{tabular}}
\end{table}

In this subsection, we discussed the different blockchain consensus methodologies which are widely adopted in practice. Each of these methods has their own set of strengths and shortcomings. A brief overview of the same is given in Table~\ref{tab1}. One major shortcoming that engulfs all the blockchain consensus algorithms is the limitation in dealing with the malicious users that exploit the features (decentralization and pseudonymity) of the blockchain technology. Among the cryptocurrency-based permissionless blockchains, no provision exists to deal with the users doing transactions attached to criminal activities except Ripple. The Ripple blockchain has worked towards restricting money laundering on its network~\cite{ripplemoney}. 

\subsection{Exploits that are prevalent in cryptocurrency-based blockchain}

Two major categories cause the threats to the credibility of a blockchain network; exploitation of vulnerabilities present and exploitation of blockchain features. The attackers exploit the vulnerabilities present to disrupt the functioning of blockchain network~\cite{LI2020841}. The presence of vulnerabilities prompts the attacks such as $51\%$ attacks, Sybil attacks, double spending, accessing the user's private keys, and exposing user identities~\cite{dasgupta2019survey}. These attacks pose harm to the network credibility by causing damage to the users such as reversal of transactions, invalidation of transactions, and user account tampering. There are a number of measures applied to prevent such attacks~\cite{FENG201945}. These measures include actions such as the release of timely patches by the developer community to mitigate the threats to the network and testing applications enabling transactions using various tools~\cite{tanmayVulnerability2021}.

Another category of threat is the threat to the blockchain ecosystem due to social engineering. The malicious entities exploit the permissionless blockchain features such as decentralization, pseudonymity, and no jurisdiction to carry out activities such as money laundering, gambling, phishing, and ransomware attacks.  The two most prominent illicit activities are ransomware transactions~\cite{8057721} and money laundering~\cite{dyntu2018cryptocurrency}. A ransomware transaction is initiated by the victim on-demand to decrypt the files on his system that an attacker encrypted. Some ransomware used by attackers include Wannacry~\cite{8260673} and CTB-Locker~\cite{7924925}. Note that ransomware is not a direct threat to the functioning of blockchain networks. However, blockchain unwittingly aids the cyber-criminal to carry out the malicious activity of extortion by providing pseudonymity. Due to the features of blockchain, it is relatively difficult for the LEAs to arrest criminals if they are outside their jurisdiction. Also, it is not easy to link blockchain users to their real-world identities without applying Know Your Customer (KYC) procedure at some cryptocurrency exchange. In money laundering, criminals sometimes perform mixing or transfer cryptocurrency having large money trails~\cite{lal2021understanding}. Apart from these, illegal betting or gambling~\cite{CONLON2020108727} is also prevalent in blockchains.

Besides above mentioned illicit activities, Silk Road (now defunct)~\cite{10.1145/2488388.2488408} was an online marketplace that exploited cryptocurrency functionalities. Within a short span of time after its creation, the silk road became a popular hub for payments for drugs trading, buying selling stolen and smuggled goods, and even contract killing. The case of the silk road was in a major way responsible for linking Bitcoin as a facilitator for criminal activities among the general populace. In the case of the silk road marketplace, the USA government successfully shut down the website as its founder was an American citizen himself. However, this has not prevented the opening of similar other services. Therefore, the identification of the users carrying out illicit activities is not enough. There is a need to implement additional functionality within the blockchain to prevent such users from exploiting blockchain functionalities.  

\section{Related Work}\label{sec:RW}

Blockchain by its structure does not differentiate between an honest and an illicit user. Therefore, the response to restrict illicit activities and the associated users is generally done after analyzing transaction data. A number of researchers have applied different ML techniques on the permissionless blockchain transaction data. The feature selection, learning, and analysis process is done off the chain after extracting the data from the blockchain.
 
\subsection{Machine learning on transaction data}

Several approaches use ML methods over cryptocurrency transaction data to detect illicit entities in cryptocurrency-based blockchains. In~\cite{harlev2018breaking} and~\cite{doi:10.1080/07421222.2018.1550550}, the authors apply supervised ML algorithms on the Bitcoin transaction data to identify the addresses linked to illicit activities such as gambling, ransomware, and illegal online marketplace. They obtain the categorized data about the illicit as well as benign activities from Chainalysis. For the data received but kept in the unidentified category, both apply different supervised ML methods, out of which gradient boosting provides the most accurate results. The features used are derived from the transaction date, amount, category of either party (sender/receiver) if defined, and transactions done with either of the predefined categories.   

In~\cite{pham2016anomaly} and~\cite{pham2016anomaly1}, the authors used the K-Means clustering algorithm for the detection of malicious users for Bitcoin transaction data. Here, the the two Bitcoin transaction data generated graphs use the clustering technique. One graph has user addresses as nodes, while the other has transactions as nodes. They train the algorithm using features such as in/out degree, unique in/out degree, average in/out transaction value, user activity duration, and user balance. The model is trained to detect potential anomalous users. The results provide detection of two out of 30 addresses in the test data. 

In~\cite{lorenz2020machine}, the authors performed active learning on Bitcoin transaction data to detect money laundering. The authors assumed to have limited ground truth about the transaction data. Active learning is applied to data clustered through unsupervised learning. The results obtained have performance similar to that of the supervised ML model with fully labeled data. The benchmark results in aforementioned case were obtained from~\cite{weber2019anti}. In~\cite{weber2019anti}, the authors contributed to the elliptic data set, claimed as the largest labeled Bitcoin transactions data set till their submission. For the prediction of illicit transactions, the Random Forest method performed best among the used supervised ML models. 

In~\cite{s20010147}, the authors proposed a system that uses ML to automate the signing process of transactions as well as detection of anomaly transactions. If a transaction is detected as anomalous, the initiator of the transaction has to sign a transaction manually. In~\cite{10.1007/978-3-030-34223-4_2}, the authors proposed a supervised ML-based fraud detection methodology for the transactions in the Ethereum blockchain. The features used in the method include average gas price, average gas limit, number of transactions, the value of transactions, and unique incoming/outgoing transactions. Among supervised ML algorithms, the authors found the Random Forest model to produce the best results. Besides these, there have been several ML-based methodologies that use features extracted from underlying temporal graphs to detect malicious accounts on Ethereum blockchain~\cite{Agarwal2021}. Here, the authors introduced the concept of bursty behavior and change in the neighborhood as features. These new features are used in addition to the features already used in other state-of-the-art methodologies. The combining of temporal graph properties and supervised ML lead to the identification of previously undetected suspicious user accounts. In~\cite{Agarwal2021}, the authors also survey different state-of-the-art approaches used within this frame. Based on their analysis, they propose time-variant probability scores following a sliding data window. Our reputation-based consensus method uses this feature for reputation values. A user's illicit behavior leads to its banishing from the consensus process for a predefined period. It locks a malicious user's account for the same period and provides an opportunity to LEAs for u=initiating action. In~\cite{lal2021understanding}, the authors detect illicit activities such as gambling, phishing, and money laundering through the identification of patterns in cyclic money transfer. Besides research-based methodologies, several companies, such as Chainalysis~\cite{chainalysis} offer services to the LEAs by analyzing the transaction data of several cryptocurrency-based blockchains.

\subsection{Decentralised reputation and trust models}

The previous section presents ML-based techniques that are used to identify if any address is involved in illicit activity. However, these techniques are usually off-chain and not integrated within the blockchain infrastructure. Besides changing the consensus mechanism to incorporate a level of security (in the case of PoS, e.g., \textit{slashing}), there are several reputation and trust models proposed to limit exploitation of blockchain vulnerabilities and infrastructure. 

In~\cite{9045608}, the authors propose a trust model based blockchain for the miners operating in the mining pool. The objective is to have a fair distribution of the reward in the pool and identification of malicious miners. In another work~\cite{9355472}, the authors propose a trust model for sharding-based blockchain networks with an objective to prevent malicious nodes from becoming shard leaders. The model computes opinion-based trust and by design works against the illicit users attempting to disrupt the blockchain. The proposed model aggregates local neighborhood opinions to compute trust values, which are often biased. In addition, the computation of these values followed by their propagation for decision-making process warrants additional network communication resources. It has a significant effect on network performance.  

Additionally, several methodologies integrate the user reputation in the blockchain consensus process with an aim to prevent illicit activities in blockchains. In~\cite{BouAbdo2020}, the authors proposed a permissionless proof of reputation consensus process. The entry for a user aspiring to join is based on a referral by an existing miner. Any new miner has to provide user wallet identity to the present network of miners and the vetting process is essential for its admission into the network. In~\cite{OLIVEIRA2020107367}, the authors proposed a reputation scheme in which randomly selected judges monitor the node behavior and update the respective reputation score accordingly. Here the judge does not monitor the context or the purpose of the transaction rather monitors the blockchain network proceedings. To limit illicit activities, here also, a new user has to be attested by at least one existing miner. A separate transaction is generated and broadcasted for this purpose. In~\cite{HuangRepchain}, the authors propose a reputation-based blockchain system with sharding. The reputation, in this case, is also associated with user (including validators) conduct (illicit conduct such as approving double-spending) within the blockchain network. Additionally, the proposed system has double-chain architecture (separate chains for transaction and reputation). The validator's reputation is decided based on transactions it approves. The shard leader finalizes the correctness of the transactions and subsequently the shard validators get their scores. In~\cite{DoDPoR}, the authors propose a ``semi-decentralized'' Delegated Proof of Reputation (DPoR) consensus based on DPoS. In this case, the user reputation score comprises of three factors; stake, personal resource utilization and transaction activity. 

In~\cite{aluko2021proof}, the authors proposed a feedback-based reputation system for blockchain networks. Their system works by having a side-chain for the reputation values and having additional communication between the users to derive normalized reputation. In~\cite{ShyamPoRF}, the authors proposed a reputation mechanism for a sharded blockchain structure with the objective to include only honest transactions. Here honest transactions refer to the ones that are not due to exploitation of any vulnerabilities present in the blockchain. Also, the miners are predecided for each shard and not randomly selected for each consensus round. The shard leaders are the users with the highest reputation score in their respective shards. In~\cite{YuRepu}, the authors proposed RepuCoin, a PoW backed reputation-based blockchain that is resilient against $51\%$ attack. The proposed structure has two types of blocks in the blockchain, keyblock for leader selection and microblock for transaction recording. The miner reputation score is based on its behavior over whether it commits transactions in accordance with the existing blockchain ledger. In~\cite{ZhuangPoR}, the authors proposed a Proof of Reputation based method where a node's reputation is based on its participation, stake, and transaction activity. The reputation value is time-variant and computed through stake and total holding time. The top $20\%$ of reputed users have most of the power concentrated amongst themselves. 

\begin{table}[htbp]
\caption{Reputation/Trust Model for Permissionless Cryptocurrency Blockchain Networks}\label{tab3}
\small{
\begin{tabular}{|l|l|l|}
\hline
\multirow{1}{*}{Ref} & \multirow{1}{*}{Features} & \multirow{1}{*}{Limitations/Drawbacks}  \\
\hline
\hline
\cite{9045608} & Peercoin blockchain, Trust model & Only relevant for informal group of miners, Extra\\
 & for miners, Protection against &  data in transaction for complaints and RepValue, \\
 &  block withholding and DDoS & Discourages individual participation by pool \\
 & &  friendly model \\
\hline
\cite{9355472} & Trust model for sharded blockchains, & Chances of bias in opinion, Additional\\
& Against illicit users, Shard selection&  communication resources for opinion values, If\\
&  based on reputation values & reputation changes, shard restructuring required \\
\hline
\cite{BouAbdo2020} & Permissionless proof of reputation, & Peer recommendation compulsory for joining,\\
& Prevention against attacks on blockchain, & Compulsory to provide wallet identity, \\
&Attacker detection using domain correlation, & User registration dependent on 3rd-party \\
&Identity theft protection  &  generated lists, Security deposit required \\
\hline
\cite{OLIVEIRA2020107367} & Judges for monitoring nodes, Judges selected & Peer recommendation compulsory for joining,\\
&  through pseudo random process & User reputation score dependent on respective\\
& Rotation based authorised miners list & judges, Separate judges' voting transactions \\
\hline
\cite{HuangRepchain} & Reputation based sharded blockchain, & Additional communication due to sidechain,\\
 & Double chain architecture, Validator& Final decision by shard leader,\\
 & reputation by approved transactions, & Separate consensus protocols for both chains,\\
 & Each shard has similar total reputation score & Reputation score based on value of transaction\\
 \hline
\cite{DoDPoR} & Delegated Proof of Reputation, & Semi-decentralized protocol,\\
 & User reputation by three factors; stake, & Long time holding of stake required, \\
 &  transaction \& personal resource utilization & Separate transaction analysis required \\
 \hline
\cite{aluko2021proof} & Feedback based reputation, Reputation of& Sidechain required, Extra communication \\
 & feedback providing users also considered, & for reputation calculation, Leader identity \\
 & Temporal scoping of reputation score &  known before start of consensus process \\
 \hline
\cite{ShyamPoRF} & Reputation mechanism for sharded blockchain & Miner nodes predecided, \\
 & Objective to include honest transactions & ``Honest'' does not include behavioural aspects,\\
 & Users with highest scores are shard leaders & User entry by shard manager's approval\\
 \hline
\cite{YuRepu} & RepuCoin, conditionally Resilient  & PoW based blockchain, \\
  & against 51\% attacks, Two types of blocks & Extra data on blockchain, \\
  & leader selection \& transaction& Reputation dependent on transactions mined \\
  & User votes dependent on reputation & \\
  \hline
\cite{ZhuangPoR} & Proof of Reputation consensus, Reputation& Dependent on stake holding time,\\
 & value is time variant, Participation Reward & Top 20\% have most power, \\
 &  proportional to reputation & Additional blocks for reputation values \\
 \hline
\end{tabular}
}
\end{table}

Table~\ref{tab3} summarizes different reputation/trust models proposed for blockchain networks. However, the aforementioned reputation-based blockchain systems have some key drawbacks. Most of them require a separate side-chain for functioning. The separate chain warrants the need for additional communication resources, which slowdowns the network performance for a large blockchain. At the same time, some proposed systems assign jury members to identify and record the behavior of the miner/validator. The functioning of jury again incurs communication resources. Also, some of the proposed systems require new users to be vetted by the existing users. Thus, making the chain pseudo-centralized, loss of anonymity, and favoritism. Finally, the state-of-the-art methods mainly focus on assigning a reputation score to the user based on the behavior that is limited to disrupting the blockchain network's functioning. They provide no comments on how we can reduce activities such as gambling or ransomware. This motivates us to answer the question the way in which the set of reputation values are used and integrated into the blockchain network, to reduce the aforementioned activities.

In Algorand~\cite{chen2019algorand} the consensus process follows the steps of byzantine agreement while giving the nodes the chance to participate in proportion to the stake they hold in the network. 
Algorand blockchain claims to solve the blockchain trilemma of decentralization, security, and scalability~\cite{blocktrilemma}. It provides deterministic block finality, and in terms of block finality time, it has quite significant results among the permissionless blockchains~\cite{8972381}. The features pertaining to the proposer/validator selection process and data encryption claim the Algorand blockchain to be most resilient against the adversarial attacks on the network among the existing permissionless blockchains. However, it has no inbuilt provision against the users engaged in illicit activities. Therefore, we incorporate and theoretically analyze our method in the Algorand blockchain consensus protocol. As we proceed, we describe our proposed methodology in detail.

\section{Proposed reputation based restriction and exclusion methodology on Algorand Blockchain}\label{sec:algo}

\begin{table}[htbp]
\caption{List of Notations.}\label{tab2}
    \resizebox{\columnwidth}{!}{\begin{tabular}{|l|l|}
\hline
Notation & Definition  \\
\hline
\hline

$N$ & Set of user accounts in the Algorand blockchain Network \\
\hline
$|N|$ & Number of user accounts in Algorand blockchain network \\
\hline
$n_{i}$ & $i^{th}$ user account in the Algorand blockchain network \\
\hline
$V_{r,1}$ & Set of users selected to propose block in step $(r,1)$ \\
\hline
$V_{r,s}$ & Set of users selected as validators for step $(r,s)$ ($s>1$)\\
\hline
$|V_{r,1}|$ & Number of user accounts selected to propose block in step $(r,1)$ \\
\hline
$|V_{r,s}|$ & Number of user accounts selected as validators for step $(r,s)$ ($s>1$)\\
\hline

$S_{n_{i}}^{r-1}$ & Stake (Algos) of user account $n_{i}$ in blockchain after finalization of $(r-1)^{th}$ block \\
\hline
$S^{r-1}$ & Total Algos available for consensus participation in round $r-1$ in algorand \\
\hline
$v_{r,s}$ & number of validator votes for step $(r,s)$ in $r^{th}$ consensus round where $s>1$\\
\hline
$Q^{r}$ & Seed value created in $r^{th}$ consensus round, used as input for VRF in $(r+1)^{th}$ consensus round. \\
\hline
$H(X)$ & Hash value output of a string $X$ \\
\hline
$SIG_{i}(H(X))$ & Value $H(X)$ signed using private key of $n_{i}$ \\
\hline
$sk^{r,s}_{i}$ & Ephemeral private key generated by $n_{i}$ for participation in step $(r,s)$ of $r^{th}$ consensus round \\
\hline
$pk^{r,s}_{i}$ & Ephemeral public key generated by $n_{i}$ for participation in step $(r,s)$ of $r^{th}$ consensus round \\
\hline
$PAY_{i}^{r}$ & Set of transactions proposed by proposer $n_{i}$ for $r^{th}$ consensus round \\
\hline
$\Phi$ & Null transaction data \\
\hline
$B^{r}_{i}$ & Block proposed by $n_{i}$ for $r^{th}$ consensus round \\
\hline
$B^{r}$ & Block finalised after $r^{th}$ consensus round \\
\hline
$n_{l^{r}}$ & User credited with successful proposing of $B^{r}$\\
\hline
$\phi^{r}$ & Empty block representation for $r^{th}$ consensus round \\
\hline
$hashlen$ & Number of bits in hash value output $H(X)$ \\
\hline
$(r,s)$ & Step number $s$ for $r^{th}$ consensus round \\
\hline
$Head(B^{r})$ & Header of finalised $r^{th}$ block \\
\hline
$T_{H}$ & Threshold number of stake units for output finalisation in each consensus step ($> 66\%$)\\
\hline
$\sigma^{r,s}_{i}$ & User credential generated by $n_{i}$ to claim participation in $r^{th}$ consensus round step $(r,s)$\\
\hline
$m^{r,s}_{i}$ & Message propagated by $n_{i}$ after participation in step $(r,s)$ \\
\hline
$\Theta^{r,1}_{i}$ & Value obtained after dividing decimal value of $n_{i}$'s credential for claim as a block proposer by $2^{hashlen}$  \\
\hline
$\Lambda^{r,s}_{j}$ & Value obtained after dividing decimal value of $\sigma^{r,s}_{i}$ in step $(r,s)$($s>1$) by $2^{hashlen}$  \\
\hline
$P^{r}_{i,v}$ & Probability of $n_{i}$ getting $v$ votes as validator for any step $(r,s)$ with $s > 1$ \\
\hline
$\Upsilon^{r,i}_{v}$ & Range for $n_{i}$'s $p^{r}_{i,v}$ value to get $v$ number of votes \\
\hline
$v^{r,s}_{i}$ & Vote propagated for step $(r,s)$ by $n_{i}$ as validator for $s > 1$ \\
\hline
$sL_{i}$ & Reputation score list generated by $n_{i}$ for all the users \\
\hline
$p^{i,j}$ & Reputation score of $n_{j}$ computed by $n_{i}$ \\
\hline
$p^{x}_{th}$ & Threshold reputation value by $n_{x}$ for other users to separate malicious and benign users \\
\hline
$ \mathcal{L}^{j,i}_{r}(s+1)$ & Loss in votes for $(r,s)$ validator $n_{i}$ as computed by $(r,s+1)$ validator $n_{j}$. \\
\hline
$\mathcal{V}^{i}_{r}(s+1)$ & Effective validator votes from $(r,s)$ computed by $(r,s+1)$ validator $n_{i}$. \\
\hline
\end{tabular}}
\end{table}

In this section, we describe the restricting methodology for the malicious user accounts in the Algorand blockchain. Here, we refer malicious user accounts as the ones using cryptocurrency transactions for either criminal activities or social engineering attacks. A user in a blockchain network can have multiple accounts with dedicated transaction history. Since our proposed method is based on transaction history, we evaluate each account against its own transaction history. For the sake of simplicity, we will henceforth refer to each user account as a separate user. We formulate and analyze the strategy for responding to users carrying out illicit activities by honest users assigned as validators. To formulate the reputation based restricting methodology, it is imperative to understand the consensus protocol of the Algorand blockchain. For the readers not familiar with Algorand blockchain consensus protocol, a brief explanation for the same is given in Appendix~\ref{app:algo}. We follow the notation convention of Algorand~\cite{chen2019algorand} with our modifications wherever required. The list of notations is summarized in Table~\ref{tab2}.

We start with the assumption that the consensus round is taking place for the $r^{th}$ block. In the first step, $(r,1)$, for consensus round of $r^{th}$ block, ($B_{m}^{r}$) is proposed by a node $n_{m} \in N$ where N is set of all the users (accounts) in Algorand. The node $n_{m}$ is selected as the proposer if $\Theta^{r,1}_{m} < \Theta^{r,1}_{x} \forall x,m \in N, x\neq m$. Here $\Theta^{r,1}_{x}$ represents normalized decimal value of the credentials of $n_x$ for 1st step. More details about the proposal selection used by Algorand's consensus process is explained in Appendix~\ref{app:algo}. Note that, for $n_{x} \in N$, in our case $x$ represents the identity of $n_{x}$ in $N$. Therefore, henceforth we use both $n_{x} \in N$ and $x \in N$ to represent the users, based on the appropriateness of either at the required place. In our manuscript, both represent the same status. 

Once the proposed blocks are broadcast into the network, the step $(r,2)$ of the consensus process starts. For steps $(r,2)$ and onward, the users selected as validators have two options. One is to accept the block $B_{m}^{r}$ as the valid addition to the blockchain, while the other option is to vote for an empty block. 

Our proposed restricting methodology first identifies whether a proposer or a validator is benign or malicious. There are a number of quantified user assessment techniques such as opinion-based or applying ML algorithms on previous blockchain transaction data described in section~\ref{sec:RW}. Our focus is to use the quantified values in a way so as to prevent illicit activities' transactions from getting added into the blockchain. Our methodology is based on a set of time-variant reputation values computed by each node, based on an agreed-upon ML model for computing the same. The reputation index is also termed as reputation score list, given as $sL_{x} = [p^{x,1} p^{x,2}, \cdots, p^{x,|N|}]$ stored by $n_{x} \in N$. Here $p^{x,y} \in [0,1]$ is the reputation value assigned by $n_{x}$ to $n_{y}$, while $n_{x}, n_{y} \in N$. As an example of computation methodology for such values, in~\cite{Agarwal2021}, the authors propose time-variant reputation scores (related to a user being malicious/benign). Here, their approach extract features from the transactions done by each user for each day and clusters the users. If a user shows behavioral similarity with known malicious users, the user gets tagged as a suspect. Over time the proposed model computes the probability of being malicious for the considered duration. The outcome of the model is same as the reputation score for our case. Note that lower the value of $p^{j,i}$, the more suspicious/malicious $n_{i}$ is as perceived by $n_{j}$. In our methodology, every user in the network is equipped with its own module to compute such reputation scores for all the users. Our method does not put any additional burden to the blockchain network communication because the input for the model is the transactions that are already replicated at each node. Such computation can be done separately from the blockchain consensus process. 

With respect to Algorand blockchain, we assume that the users compute the required values based on previous transaction data before the start of step $(r,1)$. Note that the reputation scores for at least $t$ time instances remain the same for our current approach. During these $t$ time instances, there can be many consensus rounds. During $(r,1)$, suppose a proposer $n_{m}$ proposes a block $B^{r}_{m}$ with best credentials. For step $(r,1)$, if $n_{m}$ is a benign proposer, it will avoid the transactions from the users having the low reputation values in the list $sL_{m}$. In such a case, the proposed block $B_{m}^{r}$ will get added into the blockchain. For the case when $n_{m}$ is malicious, the response of subsequent steps ($(r,2)$ and beyond) validators is discussed below.  

In step $(r,2)$, if the block proposer, as well as the validator, are benign, the validator accepts $B_{m}^{r}$ and votes in its favor. If the block proposer is benign while the validator is malicious, the validator might vote for an empty block or $B_{m}^{r}$ depending on its personal objectives. For the scenario where the block proposer is malicious while the validator is benign, the validator will vote for an empty block if the validator is aware that the proposer is malicious. For the case where both the proposer and the validator are malicious, the validator decision can go either way, depending upon its personal objectives. Note that, when a particular validator is malicious, its votes will be reduced by the other validators in the subsequent validation steps.

More formally, in step $(r,2)$ for $r^{th}$ block consensus round, suppose a validator has the identity $ n^{r,2}_{x}$, $\forall x \in V_{r,2} \subset N$, where $V_{r,s}$ is the set of users selected as validators in step $(r,s)$ for $s>1$ and $s\in \mathbb{N}$. Let $n^{r,2}_{x} = n_{j}$, $\forall n_{j} \in V_{r,2} \subset N$ with its reputation scores list given as $sL_{j} = [p^{j,1} p^{j,2} \cdots p^{j,|N|}]$. In step $(r,2)$ $n_{j}$ receives $\{Head(B^{r}_{i}), \sigma^{r,1}_{i}\}$ from $n_{i} \in V_{r,1}$ (a block proposer) after step $(r,1)$ and evaluates the normalised decimal value $\Theta^{r,1}_{i}$ of $\sigma^{r,1}_{i}$. Note that here $\sigma^{r,1}_{i}$ is the credential of $n_i$ as a proposer. Using the values $\Theta^{r,1}_{i}$ and $sL_{j}$, $n_{j}$ selects $n_{i}$ as the lead block proposer for $r^{th}$ block if $\frac{\Theta^{r,1}_{i}}{p^{j,i}} < \frac{\Theta^{r,1}_{x}}{p^{j,x}}$ for $\forall n_{x} \in V_{r,1} \subset N$ with $n_{i} \neq n_{x}$. Here $V_{r,1}$ is set of users selected as block proposer in step (r,1). 

As we mentioned earlier, our focus is on the malicious behaviour that does not attack the network directly but gradually erodes its credibility by carrying out illegal activities through malicious transactions. By use of the above methodology, the honest validators can replace a malicious proposer with an honest one without the need to push an empty block. Suppose a user $n_{m}$ has exhibited malicious behavior previously by engaging in illegal activities using blockchain network transactions. For step $(r,1)$, $n_{m}$ sends block proposal with $PAY^{r}_{m}$ consisting of suspicious transactions. In step $(r,2)$ the selected validator $n_{j}$, which considers $n_{m}$ as malicious, evaluates the proposals received and the result of the evaluation is $\Theta^{r,1}_{m} < \Theta^{r,1}_{x}$ with $\forall x \in N$ and $m \neq x$. In such a case, $n_{j}$ has to either accept the proposal of $n_{m}$ as the latter has not broken any consensus rules or vote for an empty block. The other validators in $(r,2)$ may or may not share the same viewpoint about $n_{j}$ on $n_{m}$. To resolve this potential conflict of views, a standard quantified version of viewpoint applied into consensus methodology can achieve the objective of keeping malicious activities at bay. 

\begin{figure}
\includegraphics[width=12.75cm]{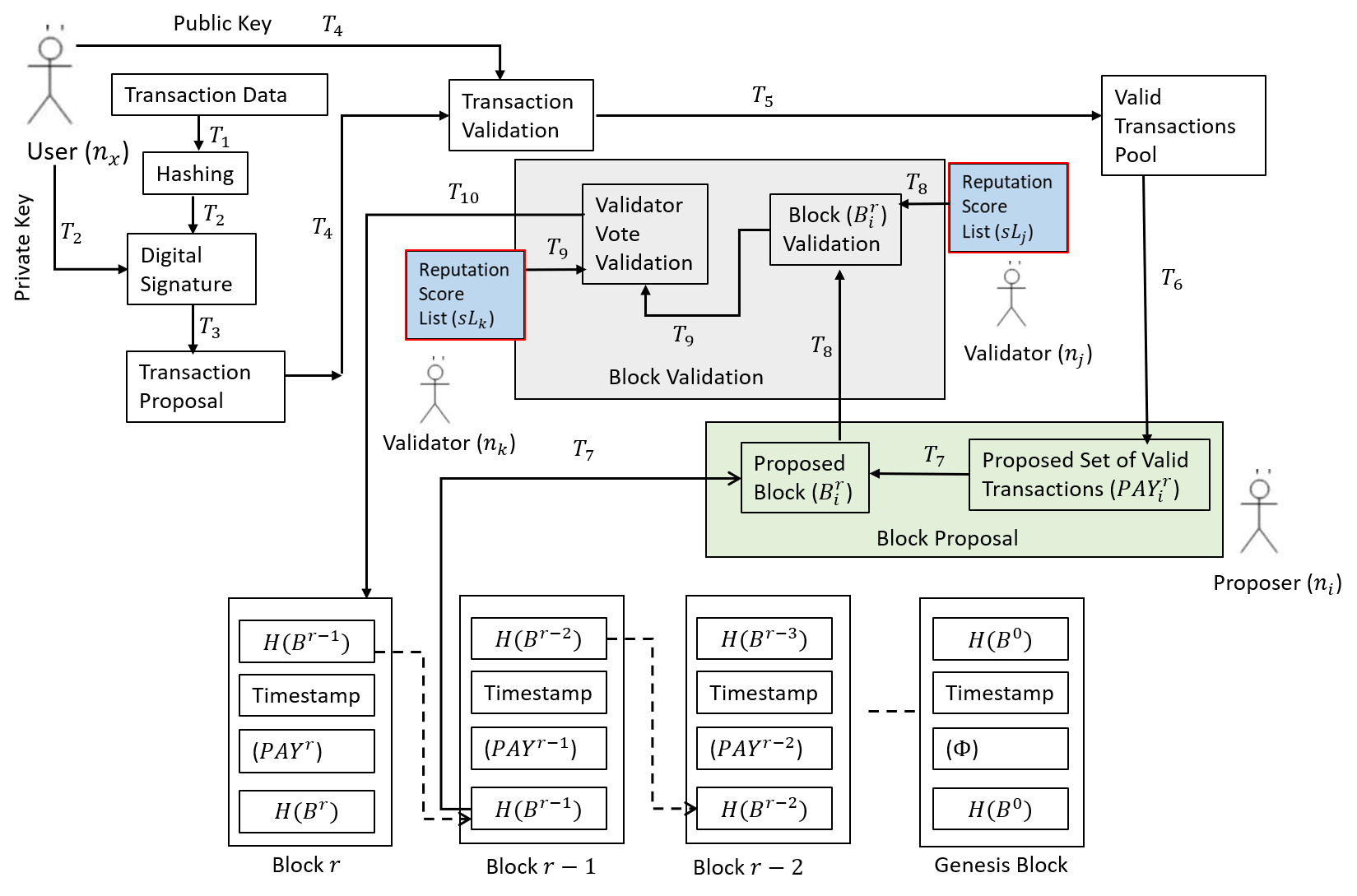}
\centering
\caption{Algorand Consensus with Reputation Score List}
\label{fig1}
\end{figure}

Figure~\ref{fig1} summarizes our modifications to the Algorand consensus process (cf. Appendix~\ref{app:algo}) and shows the parts where the reputation score list $sL_{x} \forall x \in N$ is applied. The reputation score list is used by the validators evaluating the block proposal as well as the ones evaluating previous validation step votes. We evaluate the impact due to our applied changes on the original Algorand consensus protocol in the form of theoretical results in the next section. 

\section{Results}

With the provision of reputation values list $sL_{j}$ for $n_{j}$, a malicious proposer can be replaced by an honest proposer if there is sufficient difference in their probabilities as shown below. Suppose there are two proposers $n_{i}$ and $n_{m}$ in step $(r,1)$. The user $n_{i}$ is termed as honest while $n_{m}$ is termed as malicious by other nodes. This means for a third node $n_{j}$, $p^{j,i} > p^{j,m}$. If $\Theta^{r,1}_{m} < \Theta^{r,1}_{i}$, then the block proposed by $n_{m}$ will have preference over the one proposed by $n_{i}$. The inequality condition can be written as in equation~\ref{eqcond} with the condition $\forall \delta_{x,y} \in \Re^{+}$, where $x \neq y$ and $x,y \in N$. 
\begin{equation} \label{eqcond}
    \Theta^{r,1}_{m} + \delta_{m,i} = \Theta^{r,1}_{i}   
\end{equation}
A validator $n_{j}$ in step $(r,2)$ can vote for the proposal of $n_{i}$ instead of $n_{m}$ if the condition in equation~\ref{ineqcond} satisfies for $p^{j,i},p^{j,m} \in sL_{j}$.
\begin{equation} \label{ineqcond}
\frac{\Theta^{r,1}_{i}}{p^{j,i}} < \frac{\Theta^{r,1}_{m}}{p^{j,m}}    
\end{equation}

Rearranging equation~\ref{ineqcond} and from equation~\ref{eqcond}, we get the expression as defined in equation~\ref{probcond}.
\begin{equation} \label{probcond}
    p^{j,i} > \frac{\Theta^{r,1}_{i}}{\Theta^{r,1}_{i} - \delta_{m,i}} p^{j,m}
\end{equation}
\begin{equation} \label{probcond1}
    p^{j,i} > \frac{1}{1 - \frac{\delta_{m,i}}{\Theta^{r,1}_{i}}} p^{j,m}
\end{equation}
i.e., if $n_j$ perceives $n_{i}$ to be more honest than $n_{m}$ by at least a factor of $\frac{1}{1 - \frac{\delta_{m,i}}{\Theta^{r,1}_{i}}} = C_{m,i}$ then $n_{j}$ will vote for the block of $n_{i}$ (honest proposer) in step $(r,2)$ instead of voting for either an empty block or the block of $n_{m}$ (malicious proposer).

The derivation of $C_{m,i}$ (defined as compensation factor) leads us to the formation of Lemma~\ref{lemma1}.
\begin{lem} \label{lemma1}
In the Algorand blockchain network with the provision of reputation values list, if an honest user and a malicious user are competing to propose their block, the validator $n_{j} \in N$ having reputation values list $sL_{j}$ will vote for the honest user's block if their perceived reputation of honest user is greater than that of the malicious user by atleast the multiple of compensation factor between the honest and malicious user.   
\end{lem}
\begin{cor}\label{cor1}
    In step $(r,1)$, the proposer $n_{i}$ is selected at step $(r,2)$ by the validator $n_{j}$, if the condition $p^{j,i} \in (C_{x,i} p^{j,x},1], \,\,\,\, \forall x \in V_{r,1} \subset N$ is satisfied.
\end{cor}

Since there are multiple rounds of validation, it is imperative to consider the reputation of validators as well. For steps beyond $(r,2)$, validators take into account the reputation score of the user selected as a previous step validator while considering its vote.

To understand the impact of reputation score on the validator votes, we analyze the interactions between the validators of two consecutive validation steps. 
Here, step $(r,s+1)$ validators $V_{r,s+1}$, validate the message and count the votes for validators $V_{r,s}$ in step $(r,s)$. 

Let $n_{j} \in V_{r,s}$ and $n_{k} \in V_{r,s+1}$ where $n_j,n_k \in N$. Here, $n_k$ computes $p^{k,j} \times\Lambda^{r,s}_{j}$, where $\Lambda^{r,s}_{j}$ is a normalized decimal value of the credentials of $n_j$ obtained from the voting message from step $(r,s)$.

Let $\Lambda^{r,s}_{k,j} = p^{k,j} \times \Lambda^{r,s}_{j}$. By definition, $\Lambda^{r,s}_{k,j}$ is the perceived normalised decimal value of $\sigma^{r,s}_{j}$ for $s > 1$ as per $sL_{k}$ of $n_{k}$. In Algorand consensus, $L^{r,j}_{v}$ represents the range for $\Lambda^{r,s}_{j}$ in order for $n_{j}$ to have $v$ number of votes as a validator in step $(r,s)$ (c.f. Appendix~\ref{app:algo}). With the use of $sL_{k}$, $n_{k}$ computes the number of votes assigned to $n_{j}$ based on the range (defined by Algorand consensus protocol) in which $\Lambda^{r,s}_{k,j}$ is present. Since $p^{k,j} \in [0,1]$, we get $\Lambda^{r,s}_{k,j} \leq \Lambda^{r,s}_{j}$. Suppose the perceived value $\Lambda^{r,s}_{k,j} \in \Upsilon^{r,j}_{k,v}$ where $\Upsilon^{r,j}_{k,v}$ is the range perceived by $n_{k}$ for $n_{j}$ based on $p^{k,j} \in sL_{k}$. Here $\Upsilon^{r,j}_{k,v}$ is given in the equation~\ref{percrange} where $P^{r}_{j,v}$ is probability of $n_j$ getting selected and having $v$ number of votes in the validator group in step $(r,s)$.

\begin{equation} \label{percrange}
    \Upsilon^{r,j}_{k,v} = \left [ \sum^{v}_{v'=0}P^{r}_{j,v'} p^{k,j}, \sum^{v+1}_{v'=0}P^{r}_{j,v'} p^{k,j} \right )
\end{equation}
Note that the range $\Upsilon^{r,j}_{k,v}$ either reduces the votes or does not change the number of validator votes. Supposes it reduces the validator votes from $v$ to $w$ for the step $(r,s)$. %Since $p^{k,j} \in [0,1]$, we have $w \leq v$.
The range for $\Upsilon^{r,j}_{w}$ without applying $sL_k$ can be written as shown in the equation~\ref{percrange1}.
\begin{equation} \label{percrange1}
    \Upsilon^{r,j}_{w} = \left [\sum^{w}_{v'=0}P^{r}_{j,v'}, \sum^{w+1}_{v'=0}P^{r}_{j,v'} \right )
\end{equation}
With $sL_k$, there is a difference between the actual votes and the effective votes. The loss of votes for a validator $n_j$ is directly proportional to the loss in the normalised value of its credentials from step $(r,s)$. Assume $\Upsilon^{r,j}_{w,min}=\sum^{w}_{v'=0}P^{r}_{j,v'}$ and $\Upsilon^{r,j}_{w,max}=\sum^{w+1}_{v'=0}P^{r}_{j,v'}$. Thus, we have equation~\ref{minmaxrange}.

\begin{equation} \label{minmaxrange}
    \Upsilon^{r,j}_{w} = \left [ \Upsilon^{r,j}_{w,min}, \Upsilon^{r,j}_{w,max} \right )
\end{equation}

The vote attenuation ($\mathcal{L}^{k,j}_{r}(s+1)$) in the above case is $v-w$. 
Along the similar lines, the attenuation in the value of validator $n_{j}$'s VRF output for step $(r,s)$ in step $(r,s+1)$ can be expressed as the equation~\ref{valatt}.
\begin{equation}\label{valatt}
    \mathfrak{L}^{k,j}_{r}(s+1) = \Lambda^{r,s}_{j} - \Lambda^{r,s}_{k,j} = \Lambda^{r,s}_{j} (1 - p^{k,j})
\end{equation}
The total attenuation from step $(r,s)$ counted by $n_{k}$ in step $(r,s+1)$ will be the sum of vote attenuation for all the $(r,s)$ validators and are expressed as $\mathcal{L}^{k}_{r}(s+1)$ in equation~\ref{totalatt}.  
\begin{equation}\label{totalatt}
    \mathcal{L}^{k}_{r}(s+1) = \sum_{x \in V_{r,s}} \mathcal{L}^{k,x}_{r}(s+1) \,\,\,\, \text{where } V_{r,s} \subset N
\end{equation}
Similarly, the total attenuation in the VRF output value as perceived by $n_{k}$ based on $sL_{k}$ is being given in the equation~\ref{totalvalatt}.
\begin{equation}\label{totalvalatt}
    \mathfrak{L}^{k}_{r}(s+1) = \sum_{x \in V_{r,s}} \Lambda^{r,s}_{x} (1 - p^{k,x}) \,\,\,\,  \text{where } V_{r,s} \subset N
\end{equation}
The loss in the votes is directly proportional to the loss in VRF output value. Since the values $\Lambda^{r,s}_{x}$ with $\forall x \in V_{r,s} \subset N$ are generated by the validators in step $(r,s)$, the step $(r,s+1)$ validator $n_{k}$ has no role in it. From the perspective of $n_{k}$ in step $(r,s+1)$, the loss in the votes is directly proportional to the perceived reputation values from the list $sL_{k}$.

In terms of reputation score, we have equation~\ref{votevalatt}.
\begin{equation}\label{votevalatt}
    \mathcal{L}^{k,x}_{r}(s+1) = \kappa^{r,s}_{x} (1-p^{k,x})
\end{equation}
Here, $\kappa^{r,s}_{x}$ is computed from the credential value associated with validator $n_{x}$ from step $(r,s)$. The validator $n_{k}\in V_{r,s+1}$ has no role in its computation. Note that from above, we have $\kappa^{r,s}_{x} = v$ and $p^{k,x} \kappa^{r,s}_{x} = w$. Overall, the effective votes from step $(r,s)$ perceived by $n_{k}$ are given by equation~\ref{effecvote}.
\begin{equation}\label{effecvote}
    \mathcal{V}^{k}_{r}(s+1) = \sum_{x \in V_{r,s}}  p^{k,x} \kappa^{r,s}_{x} \,\,\,\, V_{r,s} \subset N
\end{equation}
Let the total number of validator votes in step $(r,s)$ be $v_{r,s}$ for $s>1$. Note that $v_{r,s}=\sum_{x \in V_{r,s}} \kappa^{r,s}_{x}$. The attenuation in the votes as perceived by $n_{k}$ for step $(r,s)$ validators is being given in equation~\ref{totlossvote}.

\begin{equation}\label{totlossvote}
    \mathcal{L}^{k}_{r}(s+1) = v_{r,s} - \sum_{x \in V_{r,s}}  p^{k,x} \kappa^{r,s}_{x}
\end{equation}

To carry out the validation process for step $(r,s)$, it is imperative to have sufficient number of validators, thereby leading to the sufficient number of validator votes.

For the set of validators $V_{r,s+1} \subset N$ in step $(r,s+1)$, the expected vote attenuation in step $(r,s+1)$ from $(r,s)$ validators is given by the equation~\ref{totlossvote1}.

\begin{equation}\label{totlossvote1}
    \mathbb{E}(\mathcal{L}_{r}(s+1)) = v_{r,s} - \frac{\sum_{y \in V_{r,s+1}} \sum_{x \in V_{r,s}}  p^{y,x} \kappa^{r,s}_{x}}{|V_{r,s+1}|}
\end{equation}
To compensate for the votes lost, the number of validators for the next round ($r+1$) could be increased by the factor given in equation~\ref{totlossvote1} for subsequent block rounds. The compensation for the lost votes is applied in the subsequent rounds as given in equation~\ref{compvotes}. 
\begin{equation}\label{compvotes}
    v_{r+1,s} = v_{r,s} + \mathbb{E}(\mathcal{L}_{r}(s+1))
\end{equation}

The use of reputation value model not only changes the effective votes, but also amplifies the ratio of honest to malicious votes. Originally, the Algorand blockchain operates with an assumption that there are at least $\frac{2}{3}$ honest proposers for step $(r,1)$ and validators for subsequent steps in each round. Hence, there are twice as many Algos (cryptocurrency of Algorand) from honest users at stake than that with malicious users for each step $(r,s)$ of round $r$ of block formation. With the addition of $sL_k$, the effective number of votes changes as shown previously. 

We now show the effective honest and malicious votes from the perspective of validator $n_{k}\in V_{r,s+1}$ for votes of validators in step $(r,s)$. The segregation of honest and malicious $(r,s)$ validators  within $V_{r,s}$ is done by $n_k\in V_{r,s+1}$ with an assumption that their behaviour is reflected in $sL_{k}$. The malicious and honest users are separated by selecting a threshold value $p^{k}_{th}\in [0,1]$ such that $p^{k,x} > p^{k}_{th}$ for $\frac{2v_{r,s}}{3}$ and $p^{k,x} \leq p^{k}_{th}$ for remaining validators. The total effective votes are given as $\mathcal{V}^{k}_{r}(s+1)$ in equation~\ref{effecvote}. Based on $p^{k}_{th}$ and $sL_{k}$, $\mathcal{V}^{k}_{r}(s+1)=\mathcal{V}^{k,H}_{r}(s+1)+\mathcal{V}^{k,M}_{r}(s+1)$ where $\mathcal{V}^{k,H}_{r}(s+1)$ and $\mathcal{V}^{k,M}_{r}(s+1)$ are the effective votes of honest and malicious validators, respectively. These quantities are formally defined by equations~\ref{honeffecvote} and~\ref{maleffecvote}, respectively. 
\begin{equation}\label{honeffecvote}
    \mathcal{V}^{k,H}_{r}(s+1) = \sum_{x\in V^{k,H}_{r,s}} p^{k,x} \kappa^{r,s}_{x}  \,\,\,\,  \text{where } V^{k,H}_{r,s} \subset V_{r,s}, \,\,p^{k,x} > p^{k}_{th}
\end{equation}
\begin{equation}\label{maleffecvote}
    \mathcal{V}^{k,M}_{r}(s+1) = \sum_{x\in V^{k,M}_{r,s}} p^{k,x} \kappa^{r,s}_{x} \,\,\,\, \text{where } V^{k,M}_{r,s} \subset V_{r,s}, \,\, p^{k,x} \leq p^{k}_{th}
\end{equation}
In the above equations, $V^{k,H}_{r,s}$ is the set of validators which are honest from the perspective of $n_{k} \in V_{r,s+1}$. While $V^{k,M}_{r,s}$ is the set of validators which are malicious from the perspective of $n_{k} \in V_{r,s+1}$. Also, $\forall n_{k}\in V_{r,s+1}$, we have $V^{k,H}_{r,s} + V^{k,M}_{r,s} = V_{r,s}$. Similarly, we compute the loss in the validator votes in both categories as equations~\ref{hontotlossvote} and~\ref{maltotlossvote}. 
\begin{equation}\label{hontotlossvote}
    \mathcal{L}^{k,H}_{r}(s+1) = \frac{2v_{r,s}}{3} - \sum_{x\in V^{k,H}_{r,s}}  p^{k,x} \kappa^{r,s}_{x}
\end{equation}
\begin{equation}\label{maltotlossvote}
    \mathcal{L}^{k,M}_{r}(s+1) = \frac{v_{r,s}}{3} - \sum_{x\in V^{k,M}_{r,s}}  p^{k,x} \kappa^{r,s}_{x}
\end{equation}
From equations~\ref{votevalatt} and~\ref{totlossvote}, we have $\sum_{x\in V^{k,H}_{r,s}} \kappa^{r,s}_{x} = \frac{2v_{r,s}}{3}$ and $\sum_{x\in V^{k,M}_{r,s}} \kappa^{r,s}_{x} = \frac{v_{r,s}}{3}$.
Hence, the relation between the honest and malicious votes is given as equation~\ref{honmalvot}.
\begin{equation}\label{honmalvot}
    \sum_{x\in V^{k,H}_{r,s}} \kappa^{r,s}_{x} = 2\sum_{x\in V^{k,M}_{r,s}} \kappa^{r,s}_{x}
\end{equation}

Based on Algorand's $\frac{2}{3}$ honest nodes assumption and $p^{k}_{th}$, we sort all the reputation scores. 

Since the highest reputation value in a malicious validator set is less than the lowest reputation value of an honest validator set, the loss in the votes for the malicous validator set will be more than the same in the honest validator set. The relation between the losses can be derived using equations~\ref{hontotlossvote} and~\ref{maltotlossvote} as equation~\ref{lossvotrel}.
\begin{equation}\label{lossvotrel}
    \mathcal{L}^{k,H}_{r}(s+1) \leq 2\mathcal{L}^{k,M}_{r}(s+1)
\end{equation}
Since the loss in votes is less in the case of honest validators set, the effective number of votes will relatively increase for the same against the malicious validator set. Therefore, 
the relation between the effective votes of honest and malicious validators for step $(r,s)$ is given as equation~\ref{effvotrel}.
\begin{equation}\label{effvotrel}
     \sum_{x\in V^{k,H}_{r,s}} p^{k,x} \kappa^{r,s}_{x} \geq 2\sum_{x\in V^{k,M}_{r,s}} p^{k,x} \kappa^{r,s}_{x}
\end{equation}

Alternatively, the above relation is written as equation~\ref{effvotrel1}.

\begin{equation}\label{effvotrel1}
    \mathcal{V}^{k,H}_{r}(s+1) \geq 2 \mathcal{V}^{k,M}_{r}(s+1)
\end{equation}

Let $\Delta \mathcal{L}^{k}_{r}(s+1)=2\mathcal{L}^{k,M}_{r}(s+1) - \mathcal{L}^{k,H}_{r}(s+1)$. Here $\Delta \mathcal{L}^{k}_{r}(s+1)\geq0$ is the parameter notifying the difference in the loss of validator votes for honest and malicious set. From above, the increase in the effective votes of honest validators in step $(r,s)$  with respect to that of the malicious validators from the perspective of validator $n_{k}\in V_{r,s+1}$ is given in the equation~\ref{cheffvot}.

\begin{equation}\label{cheffvot}
    \mathcal{V}^{k,H}_{r}(s+1) = 2 \mathcal{V}^{k,M}_{r}(s+1) + \Delta \mathcal{L}^{k}_{r}(s+1)
\end{equation}
The ratio of honest to malicious validator votes is henceforth given by the equation~\ref{cheffvot1}.
\begin{equation}\label{cheffvot1}
    \frac{\mathcal{V}^{k,H}_{r}(s+1)}{\mathcal{V}^{k,M}_{r}(s+1)} = 2  + \frac{\Delta \mathcal{L}^{k}_{r}(s+1)}{\mathcal{V}^{k,M}_{r}(s+1)} = 2  + \Hat{L}^{k}_{r}(s+1) 
\end{equation}

Note that $\frac{\Delta \mathcal{L}^{k}_{r}(s+1)}{\mathcal{V}^{k,M}_{r}(s+1)}$ is the relative difference in the loss of votes with respect to the malicious validator votes as perceived by $n_{k}\in V_{r,s+1}$. Let the aforementioned term be represented as $\Hat{L}^{k}_{r}(s+1)$. 
 
The expression in the equation~\ref{cheffvot1} leads us to the formation of Lemma~\ref{Lemma2}.

\begin{lem} \label{Lemma2}
In the Algorand blockchain  with the provision of reputation values list, the ratio of honest to malicious validator votes of step $(r,s)$ for the formation of $r^{th}$ block increases by the value $\Hat{L}^{k}_{r}(s+1)$ above the originally assumed factor of $2$ as evaluated by a step $(r,s+1)$ validator $n_{k} \in V_{r,s+1}$ using its reputation values list. 
\end{lem}

On average (expected value) of relative difference in the loss of votes with respect to the malicious validator votes as perceived in step $s+1$ is given by $\mathbb{E}(\Hat{L}_{r}(s+1)) =  \frac{\sum_{x\in V_{r,s+1}}\Hat{L}^{x}_{r}(s+1)}{|V_{r,s+1}|}$.

This led us to formulate corollary~\ref{cor2} from the Lemma~\ref{Lemma2}. 
\begin{cor}\label{cor2}
    In the Algorand blockchain with the provision of reputation values list, the expected ratio of honest to malicious validator votes of step $(r,s)$ for the formation of $r^{th}$ block increases by the value $\mathbb{E}(\Hat{L}_{r}(s+1))$ above the originally assumed factor of $2$ as perceived in step $(r,s+1)$ for $s > 1$.
\end{cor}

The improvement in the ratio of honest to total validator votes is beneficial in the case when the original ratio falls below $\frac{2}{3}$. As given in corollary~\ref{cor2}, the Algorand blockchain can tolerate a drop of $\mathbb{E}(\Hat{L}_{r}(s+1))$ from the ratio of 2 for honest to malicious votes. For such an amount of drop, it can be compensated and reached to the required honest majority.  

Again, the Algorand blockchain assumes that there are $\frac{2}{3}$ honest users/cryptocurrency-units selected in each step of the consensus round, including block proposal and validation. Hence, while proposing the block in step $(r,1)$, the probability of a malicious proposer getting selected as the highest priority proposer is $\frac{1}{3}$. One way in which a malicious proposer can harm the blockchain is by proposing a block consisting of transactions associated with malicious activities.  
 
Suppose in step $(r,2)$, a set of validators $n_{j} \in  V_{r,2} \subset N$ are selected to asses step $(r,1)$ block proposals from $n_{i} \in  V_{r,1} \subset N$. In step $(r,2)$, if an honest validator receives this proposed block from a malicious proposer ($n_i$) with lowest $\Theta^{r,1}_i$, he has two choices, either vote for the same or to propagate the vote for an empty block. In the absence of a defined mechanism to identify the malicious proposer, an honest validator may well be unaware and go with the first choice out of the aforementioned ones. In case the validator is aware, it will push for the empty block as there is no mechanism defined to identify the second-best credential among the block proposers. 
  
Let the probability that $n_{j}$ votes for empty block ($\phi^{r}$) be $\beta^{r}_{j,\phi}\in [0,1]$, while the same for voting $n_{i}$'s block proposal be $\beta^{r}_{j,i}\in [0,1]$. We now discuss the scenarios with validators when the block proposer is malevolent and pushing malicious transactions without and with reputation values list.  

\begin{enumerate}
    \item \textbf{$n_{j} \in V_{r,2}$ is honest and aware that $n_{i} \in V_{r,1}$ is malicious:}

In the case without reputation values list, $\beta^{r}_{j,\phi} >> \beta^{r}_{j,i}$ with $\beta^{r}_{j,\phi} + \beta^{r}_{j,i} = 1$. Note that, for the block by an honest proposer $n_{k}\in V_{r,1}$ where $\Theta^{r,1}_{i} < \Theta^{r,1}_{k}$, we have $\beta^{r}_{j,k} = 0$. 

For the case where $n_{j}$ has the reputation values list $sL_{j}$, $n_{j}$ will have the value $p^{j,i}$ related to $n_{i}$. To derive the condition for $n_{j}$ not selecting empty block, we introduce the concept of pseudo-sublists. A  pseudo-sublist, for a step $(r,2)$ validator $n_{j}$ is $sL^{r,2}_{j}\subset sL_{j}$ and consists $p^{j,x}$ where $x \in V_{r,1}$. 

In line with the Algorand's assumption of $\frac{2}{3}$ honest proposers and $\frac{1}{3}$ malicious proposers, $p^{j}_{th}$ is chosen so as to divide $sL^{r,2}_{j}$ into the same. For $n_{i}\in V^{M}_{r,1} \subset V_{r,1}$ and $n_{k}\in V^{H}_{r,1} \subset V_{r,1}$, we have $p^{j,k} > p^{j}_{th}\geq p^{j,i}$ derived from equations~\ref{honeffecvote} and~\ref{maleffecvote}. Now, the validator $n_{j}$ will not vote for an empty block if the most qualified block proposer is honest. From Lemma~\ref{lemma1}, we know that block proposal of $n_{k}$ will be chosen over that of $n_{i}$ by $n_{j}$ if $p^{j,k} > C_{i,k}p^{j,i}$. The condition to obtain $\beta^{r}_{j,\phi}=0$ and $\beta^{r}_{j,i}=0$ with absolute surety is $p^{j,k} > C_{i,k}p^{j}_{th}$
where $n_{i}$ is a malicious proposer as perceived by $n_{j}$, i.e., $p^{j,i} \leq p^{j}_{th}$ with the highest proposer credentials (least $\Theta^{r,1}$ value). The aforementioned outcome can be summarised in the Corollary~\ref{cor3}.
\begin{cor}\label{cor3}
    In the Algorand blockchain, an honest and aware validator, when provided with a reputation values list, will reject an empty block with absolute surety, if there is an honest proposer present with it's perceived reputation value greater than the product of threshold reputation value and relative compensation factor relative to malicious proposer with highest credentials.  
\end{cor}
    \item \textbf{$n_{j}$ is honest but unaware that $n_i$ is malicious and thinks $n_i$ is honest:}
    
Without the reputation values list, if the validator $n_{j}$ is honest but unaware of the $n_{i}$ being malicious, it will most likely vote for the $n_{i}$'s block proposal.     
However, for an honest but unaware validator $n_{j}$ without reputation values list, $\beta^{r}_{j,\phi} << \beta^{r}_{j,i}$ with $\beta^{r}_{j,\phi} + \beta^{r}_{j,i} = 1$. Also, for a proposed block of honest proposer $n_{k}\in v_1$ with $\Theta^{r,1}_{i} < \Theta^{r,1}_{k}$, we have $\beta^{r}_{j,k} = 0$.

With the reputation values list $sL_{j}$, the behavior of $n_{j}$ will be the same as the case of being honest and aware. However, the role of the reputation values list now is to prevent $n_{j}$ majorly from voting for a malicious proposer's block. The condition to obtain $\beta^{r}_{j,\phi} = 0$ and $\beta^{r}_{j,i} = 0$ will be same as in previous case and can be summarised in Corollary~\ref{cor4}.
\begin{cor}\label{cor4}
    In the Algorand blockchain, an honest but unaware validator, when provided with a reputation values list, will reject a malicious proposer's block with absolute surety, if there is an honest proposer present with it's perceived reputation value greater than the product of threshold reputation value and relative compensation factor relative to malicious proposer with highest credentials.  
\end{cor}
\end{enumerate}
In this section, we discussed the cases where the validator is honest, which is equivalent to the assumption that they operate according to the blockchain consensus methodology. For the cases where the validators are themselves malicious, their response to malicious proposers depends on the factor of whether they are operating individually or in a group. Since their behavior is not in accordance with the following of everything according to blockchain consensus methodology, the question of using a reputation values list does not arise for them. The objective of the reputation values list is to strengthen the power of honest users in the network and provide them with the tool to prevent malicious transactions from getting included in the blockchain. The malicious users do not consider the reputation values while making decisions as either proposers or validators. Therefore, one cannot say with absolute surety whether a malicious validator will reject a malicious proposer's block, reject an empty block or select either of both. 

The validator $n_{j}$ in step $(r,2)$ follows the condition in equation~\ref{ineqcond} to assess the block proposals from step $(r,1)$. Now, if a malicious user is set to become the proposer with best suited credentials, it can be replaced by an honest user's block by $n_{j}$ if the condition in Lemma~\ref{lemma1} is satisfied. Now, in step $(r,1)$, if $n_{i}$ is a malicious proposer while $n_{k}$ is an honest proposer, then the block of $n_{k}$ must be accepted by more than $\frac{2}{3}$ of step $(r,2)$ validators (money units) based on Lemma~\ref{lemma1} in order for it to move towards getting finalised.

\section{Analysis}\label{sec:analysis}

We evaluate our methodology over the parameters such as completeness, soundness, communication hop, and extra steps in communication. In~\cite{chen2019algorand}, authors provide the analysis of the Algorand blockchain network consensus mechanism with respect to proof of the aforementioned properties. As we proceed, we check whether the use of the reputation scores list violates any of the proved properties of the Algorand blockchain network. If that is not the case, then the modified consensus mechanism also upholds all the original properties.

According to Theorem $1$ in~\cite{chen2019algorand}, there are $4$ properties which hold with ``overwhelming probability'' for each consensus round $r \geq 0$. Property 1 states that ``all honest users agree on the same block $B^{r}$, and all payments in $B^{r}$ are valid''~\cite{chen2019algorand}. In the case where the honest users have the reputation values list, the values in the list get generated from the blockchain transactions data, which is the same across the network. Also, the reputation values are generated and updated at regular intervals based on previous data, and their generation is not linked to the communication in block consensus rounds. All the honest users follow the methodology according to equation~\ref{ineqcond} and agree on the same block. 

The property $2$ of Theorem $1$ in~\cite{chen2019algorand} states that if the consensus leader $n_{l^{r}}$ is honest, then the generated block $B^{r}$ is known to the honest users in a predefined time interval. Here, the consensus leader refers to the block proposer $n_{l^{r}}$ in step $(r,1)$ such that $\Theta^{r,1}_{l^{r}} < \Theta^{r,1}_{x}$ for $l^{r}, x \in N$. It also specifies the amount of time by which the finalized block $B^{r}$ will be known to the first honest user for the cases when payment set $PAY^{r}$ is empty and non-empty. The proposed modification does not involve any change in the communication procedure. The changes in the decision-making method, as shown in equation~\ref{ineqcond} is done at each user's end only. Also, there is no change proposed in the rules regarding the finalization of a block. Therefore, the property $2$ also upholds under the proposed modifications. 

The property $3$ discusses about the time taken in block finalization when $n_{l^{r}}$ is malicious. Under the proposed modifications, if $n_{l^{r}}$ is declared malicious by the reputation values list, its proposal gets rejected if there is an honest proposer satisfying the condition in Lemma~\ref{lemma1} as evaluated by the honest users. If not, then the process takes place as in the original Algorand blockchain network. Since the evaluation is done at the user's end and not within the communication, the block finalisation happens within the predefined time interval only. 

The property $4$ of Theorem $1$ in~\cite{chen2019algorand} states that the probability of $n_{l^{r}}$ being honest is $p_{h} = h^{2}(1+h-h^{2})$, where $h$ is the ratio of honest user Algos at stake in the network. From Lemma~\ref{Lemma2}, we know that the ratio of honest to malicious user money units increases for the consensus steps involving block validation if the reputation values list is used for evaluation. Therefore, the value of $p_{h}$ also increases in such a case. In conclusion, all the $4$ properties of Theorem $1$ in~\cite{chen2019algorand} uphold even with the proposed modifications in the consensus procedure using the reputation values list. 

\subsection{Change in time taken}
The process to compute the reputation values list is done at the user's end without attachment to the consensus process. When a user participates as either a proposer or a validator, the reputation values are already available pre-computed. While using these values, the evaluation process for selecting a block and counting the votes remain same. Therefore, there is no change in the time taken during the consensus procedure. 

\subsection{Completeness}
According to the completeness lemma in~\cite{chen2019algorand}, if the properties $1-3$ in Theorem $1$ hold for consensus rounds $\{0,1,\cdots,r-1\}$, then properties $1$ and $2$ hold for consensus round $r$, when the consensus leader $n_{l^{r}}$ is honest. As we have discussed above, the use of the reputation values list does not have any effect on the upholding of either of the aforementioned properties. Hence the completeness will uphold for the modified consensus procedure also. 

\subsection{Soundness}
According to the soundness lemma in~\cite{chen2019algorand}, if the properties $1-3$ in Theorem $1$ hold for consensus rounds $\{0,1,\cdots,r-1\}$, then properties $1$ and $3$ hold for consensus round $r$, when the consensus leader $n_{l^{r}}$ is malicious. In the case of $n_{l^{r}}$ being malicious, its proposal will be replaced by an honest users' proposal if it satisfies the condition defined in Lemma~\ref{lemma1} as evaluated by the honest users. Even if that is not the case, the consensus process will happen according to the original methodology. Therefore, the soundness of the network will remain intact even with the proposed modifications. 

\subsection{Change in local space storage and communication overhead}
The use of a machine learning algorithm to calculate reputation values warrants some additional space at an individual user's end. However, as mentioned earlier, the process is not tied up to the communication part of the consensus mechanism and will not result in the addition of any overhead in communication. 

\subsection{Change in communication hop}
Our proposed methodology involves evaluation on the users' end only; the communication data will remain the same as in the original Algorand blockchain network. Therefore, there will be no change in the communication hop.

\section{Discussion and Conclusion}\label{sec:discussion}

In this paper, we propose a reputation-based methodology for applying the reputation values in the Algorand blockchain to restrict illicit activities by criminals who are using the platform. Our reputation method is based on the behavior of an account in terms of transactions. It assigns a reputation score $\in [0,1]$ to each account where the level of honesty increases from $0$ to $1$. For obtaining the reputation scores through transaction data, different ML models are used. A time-variant reputation score based on the ML model operating through sliding window input data is ideal input for our proposed methodology. As observed in related work on ML models, a 24 hour window for updating scores is sufficient to provide insight into the overall network behavior. Also, it was observed in the related work that there is not much deviation in the user behavior observed in the bitcoin and Ethereum blockchain. It provides us with a starting point in terms of the choice of ML model to provide input for our methodology. The identification of reputation score leverages the distributed nature of the blockchain technology, where each account can identify the reputation score for all the accounts. Thus it aids in identifying suspects operating within the blockchain ecosystem. Our method involves minor changes in the consensus protocol at the user end. Locally, in the original consensus protocol, all user accounts evaluate at their end and identify whether that particular account is a proposer, validator, or both. We do not change the communicated messages; rather we modify the evaluation for being a proposer, a validator, or both. The other steps in the consensus protocol remain the same as before. The reputation score plays a major part in identification of the proposer/validator.
Our methodology involves eliminating blocks proposed by suspicious accounts and reducing the validation votes for the suspicious validators. Also, our method, although designed for user behaviour based reputation, can also accommodate the adversarial attacks within the reputation score. Overall, our methodology aims to enhance the decision-making process, have maximum possible impact, and cause minimum possible disruption at the communication level. 

However, our proposed methodology is not without certain limitations. The output of the reputation score is obtained by applying the pre-decided ML algorithm on the transaction data present on the respective blockchain. Although the input data and the features used are the same, there is a chance of variation in the output across different users running algorithm at their end. Therefore, the selection of a proper ML algorithm with rigorous cross-platform testing is a way to avoid the scenario mentioned above and its effects such as exclusion of benign users and selection of empty block due to divided opinion. Additionally, the users have to run the ML algorithm at their end. Although it does not burden communication resources, the user has to perform additional computational tasks out of network for decision-making according to the proposed methodology.   

Our approach is achievable for other blockchain networks as well, with the mechanism tuned to the respective consensus methodologies. Ethereum blockchain network is transitioning from PoW to PoS consensus~\cite{ethereumtransition}. The transition is supposed to be through enabling the use of Casper~\cite{buterin2017casper}, a hybrid PoW/PoS protocol. Here, the PoS is applied to finalize the blocks after a fixed number of blocks are mined through PoW. This change aims to reduce the chances of illicit activities such as invalidating the blocks, forming a parallel chain, and disrupting the communication by forfeiting the deposit of the validators. However, still, there is no provision to stop social-engineering based illicit activities such as Phishing, money laundering, Gambling. Applying the reputation model based on transaction data paves the way for identifying the entities using blockchain for the aforementioned illicit activities. The identification helps in curtailing the power of the proposer and validator accounts involved in such social-engineering based illicit activities within the blockchain network. 

For complete PoW based blockchain networks such as Bitcoin, our reputation model can play an advisory role to prevent transactions from illicit entities from getting recognized. Since mining a single block is computationally expensive, an honest miner would not risk its block getting rejected by including the transactions from illicit entities. Since the scores get updated with the latest behavior playing a more significant role, the reputation scores can effectively freeze a malicious user account and give the law enforcement agencies enough time to catch up with the same. 

Similarly, our reputation score based regulation is also relevant for the directed acyclic graph (DAG) based blockchain such as IOTA Tangle~\cite{SILVANO2020307}. In Tangle, a new transaction selects previous transactions based on Markov Chain Monte Carlo algorithms~\cite{geyer1992practical} to verify and join the DAG. By using our reputation scores, the selection procedure can give a transaction a reputation score based on the user behavior. An honest user wants to operate according to the blockchain network rules and associate with fellow honest users. It also wants to play its part in keeping the blockchain network free from illicit entities. Hence, the reputation score can be used by the new transaction to select those transactions to join with which are done by the honest users.

Further, in extended proof of stake such as delegated proof of stake (DPoS) used in Cosmos~\cite{kwon2019cosmos} our methodology is also applicable. In DPoS, the non-validators can delegate their cryptocurrency to validators for a share in block fees. The users in DPoS based blockchain can use the reputation score model to identify the validators suitable for delegating their cryptocurrency stake. Therefore, the concept of restricting illicit activities from exploiting blockchain infrastructure is also applicable in this case. Note that here we provide the theoretical proof for Algorand while a strong case for other blockchains to incorporate our proposal.

This paper aims to provide a theoretical foundation and do the feasibility analysis of the regulation on the blockchain network against criminal entities. The idea behind the conception of blockchain technology was the establishment of a self-regulated decentralized system. Our proposed methodology provides the means to move further towards achieving the same goal. A blockchain user, when equipped with the reputation score, can make the decision on whether to carry out the transaction with the user perceived as malicious based on reputation score values. For a consensus round, a potential block proposer has the choice to include the transactions it perceived as honest. Therefore, an illicit activity related transaction will remain in pending state and eventually expire after end of its validity period. It will promote honest behavior across the blockchain network. In the discussion of the proposed methodology and its analysis, we show that there is no effect on the communication pattern and data, thereby no additional strain on the network resources. The theoretical results obtained demonstrate that an honest validating committee member can prevent an identified criminal entity from proposing the block on the Algorand blockchain. In addition, exploiting the blockchain transactions for illicit activities restricts the respective user's role as a validator. Overall, we conclude that the proposed consensus-based regulation methodology is feasible enough to be integrated within the blockchain ecosystem.  

\section*{Acknowledgement}
This work is partially funded by the National Blockchain Project (grant number NCSC/CS/2017518) at IIT Kanpur sponsored by the National Cyber Security Coordinator's office of the Government of India and partially by the C3i Center funding from the Science and Engineering Research Board of the Government of India (grant number SERB/CS/2016466).

 \bibliographystyle{plain}
  \bibliography{biblio}
  
\appendices

\section{Algorand consensus process} \label{app:algo}

In Algorand~\cite{chen2019algorand}, the consensus process follows a hybrid PoS-BFT mechanism. As we proceed, we describe the process steps leading to block formation and addition into the blockchain. Our assumption is that the current blockchain already contains $r-1$ blocks, where $r \geq 3$ and $r \in \mathbf{N}$. Therefore, the consensus process is taking place for $r^{th}$ block. The notations used are based on \cite{chen2019algorand}. In Algorand, the users in the blockchain network are in the set $N$ with the stake of user $n_{i} \in N$ till $(r-1)^{th}$ block given as $S^{r-1}_{n_{i}}$.  The consensus steps for $r^{th}$ block formation round are termed as $(r,s)$, where $s \in \mathbf{N}$ and represents the step number. Step $(r,1)$ is the block proposal step, where the selected nodes create and propagate a block into the network. To propose the $r^{th}$ block, an essential component is a seed $Q^{r-1}$ defined  using equation~\ref{algoseed}.

\begin{equation} \label{algoseed}
 Q^{r-1} = \left\{\begin{matrix}
  H(SIG_{l^{r-1}}(Q^{r-2}, r-1)) & if \; B^{r-1} \neq \phi^{r-1}\\ 
 H(Q^{r-2}, r-1) & if \; B^{r-1} = \phi^{r-1} 
\end{matrix}\right.
\end{equation}

Here, $\phi^{r-1}$ represents empty $(r-1)^{th}$ block, which is the case when none of the proposed blocks is able to secure required consensus for round $r-1$. Also, the term $H(SIG_{i}(Z))$ is defined as the hash value of the quantity $Z$ after being digitally signed by user $n_{i}$ using its private key, $sk^{r,1}_{i}$, generated for the same purpose. An empty block is defined as in equation~\ref{empblock} where $\Phi$ represents `NULL' transaction data.

\begin{equation} \label{empblock}
 \phi^{r-1} = (r-1, Q^{r-2}, H(B^{r-2}), \Phi)
\end{equation}

In equation~\ref{algoseed}, the value of $Q^{r-1}$ in case of $B^{r-1}$ being non-empty is calculated by the node $l^{r-1}$. Here $l^{r-1} \in N$ is the successful proposer credited with addition of confirmed block $B^{r-1}$ at round $r-1$. The value of $Q^{r-1}$ is determined with the finalisation of block $B^{r-1}$. In the process of formation of $B^{r}$, $Q^{r-1}$ is used to determine the proposers and the validators for $B^{r}$. For participating in consensus process for $r^{th}$ block, a node $n_{i}$ computes hash value $H(SIG_{i}(r,s,Q^{r-1}))$ in each step $(r,s)$ for $s>1$. The hash value calculation for $s=1$ is done while considering the user's stake in the network. 

The selection procedure for the proposers and validators is known as cryptographic sortition. In step $(r,1)$, each node (user account) $n_{i} \in N$ computes its credential $\sigma^{r,1}_{i}$, and checks the output against a predefined threshold. The procedure for consensus is effectively byzantine agreement combined with the proof of stake. To incorporate the advantage based on the stake in the network, for step $(r,1)$, each unit of Algorand cryptocurrency is treated as an individual sub-user attached to the user holding it. Therefore, the more a user's stake is, the more it has sub-users, and the better its chances are of getting selected as a block proposer. The credential which $n_{i} \in N$ generates for step $(r,1)$ and propagates in case of getting selected as proposer is given in equation~\ref{propcred}. The proposer $n_{i} \in V_{r,1}$ selects $\sigma^{r,1}_{i}$ such that $H(\sigma^{r,1}_{i})$ is minimum $\forall K \in [1, \cdots, S^{r-1}_{n_{i}}]$. 

\begin{equation} \label{propcred}
    \sigma^{r,1}_{i} =  SIG_{i}(r,1,K, Q^{r-1})  
\end{equation}

For step $(r,1)$, the hash value of the credential is converted into decimal and then normalised using division by $2^{hashlen}$, where $hashlen$ is the length of the output hash $\sigma^{r,1}_{i}$ for $n_{i}$. Let the normalised decimal value of $\sigma^{r,1}_{i}$ be $\Theta^{r,1}_{i}$. To consider the stake for $n_{i}$ in step $(r,1)$, the hash value to compute will be $H(SIG_{i}(r,1,K, Q^{r-1}))$, where $K \in [1,S^{r-1}_{n_{i}}]$ represents the $K^{th}$ sub-user alias of $n_{i}$.

If $n_{i}$ gets selected in $(r,1)$ as a proposer, it forms a block $B^{r}_{i}$ to propagate it to its peers in the blockchain network. Here, $B^{r}_{i}$ is the proposed block by $n_{i}$ which can get confirmed as $B^{r}$ after successfully getting majority validation in the consensus process (explained next). The cutoff number for majority validation is also predefined in the network and is referred to as $T_{H}$. The components of non-empty $B^{r}_{i}$ are given in equation~\ref{propblock}, where $Head(B^{r}_{i})= \{r, SIG_{i}(Q^{r-1}), H(B^{r-1})\}$ is the header of the proposed block.

\begin{equation} \label{propblock}
     B^{r}_{i} = \{r, SIG_{i}(Q^{r-1}), H(B^{r-1}), PAY^{r}_{i}\}
 \end{equation}

Here, $PAY^{r}_{i}$ is the set of transactions to be included in $r^{th}$ block as proposed by $n_{i}$, while $H(B^{r-1})$ is the hash value of the previous block. After forming $B^{r}_{i}$, $n_{i}$ propagates its block proposal in the form of message, $m^{r,1}_{i}$, along with a lightweight message $\{Head(B^{r}_{i}), \sigma^{r,1}_{i}\}$ confirming its selection in $(r,1)$ round. The lightweight message propagates across the network and reaches other users faster than $m^{r,1}_{i}$. It makes the evaluation of the block proposal faster. The message $m^{r,1}_{i}$ contains $\{B^{r}_{i}, SIG_{i}(H(B^{r}_{i})), \sigma^{r,1}_{i}\}$ (full block data and its credentials). To get selected as the block proposer, $n_{i}$ has to satisfy the condition $\Theta^{r,1}_{i} < \Psi^{r,1}$. Here $\Psi^{r,1}$ is a predefined threshold value for selection in first round. 

For $s>1$, the user credential generated are $\sigma^{r,s}_{i} = SIG_{j}(r,s,Q^{r-1})$ for $n_{j} \in N$. If $n_{j}$ gets selected as validator in step $(r,s)$ for $s>1$, then $n_{j} \in V_{r,s}$. Here $SIG_{j}(r,s,Q^{r-1})$ is the encrypted value of the string $(r,s,Q^{r-1})$ signed using $n_{j}$'s private key ($sk^{r,s}_{j}$) from its ephemeral public-private key pair $\{pk^{r,s}_{j}, sk^{r,s}_{j}\}$. The hash value of the credential $\sigma^{r,s}_{j}$ is used by $n_{j}$ in round $(r,s)$ to check for two outcomes, whether it is selected and if selected, how many votes as validator it got assigned for $(r,s)$ step. Let $n_{j}$ have the stake $S^{r-1}_{j}$ till $(r-1)^{th}$ block, with total network stake being $S^{r-1}$. Note that the record of the stake is verified by blockchain data. Let the set of users  to be selected as validators in step $(r,s)$ be defined by $V_{r,s}$. In such case, the probability for selection of an individual money unit as validator in $(r,s)$ step for $s>1$ is given as $P^{r}_{j} = \frac{S^{r-1}_{j}}{S^{r-1}}$. Therefore, the probability of $n_{j}$  getting selected and having $v$ number of votes in the validator group in step $(r,s)$ for $s>1$ is given in equation~\ref{valprob}.    

\begin{equation} \label{valprob}
    P^{r}_{j,v} = \binom{S^{r-1}_{j}}{v} (P^{r}_{j})^{v} (1 - P^{r}_{j})^{S^{r-1}_{j} - v}
\end{equation}

We also have $\sum^{S^{r-1}_{j}}_{v=0}P^{r}_{j,v} = 1$. Now, the range $[0,1]$ is divided into $S^{r-1}_{j} + 1$ sub-ranges. The length of $v^{th}$ sub-range for $n_{j}$, denoted by $\Upsilon^{r,j}_{v}$ is given in equation~\ref{valrange}.

\begin{equation} \label{valrange}
    \Upsilon^{r,j}_{v} = \left [\sum^{v}_{v'=0}P^{r}_{j,v'}, \sum^{v+1}_{v'=0}P^{r}_{j,v'} \right )
\end{equation}

In the consensus process, the assumption is that at least $\frac{2}{3}$ of the stake is with honest users. The threshold for confirming the validation result in each step $(r,s)$ of round $r$ for $s>1$ is $t_{H} = \frac{2|V_{r,s}|}{3}+1$. Let the normalised decimal value of $\sigma^{r,s}_{j}$ for $s > 1$ be $\Lambda^{r,s}_{j}$. In order for $n_{j}$ to get selected and have $v$ number of votes as validator in step $(r,s)$ of round $r$ for $s > 1$, the condition $\Lambda^{r,s}_{j} \in \Upsilon^{r,j}_{v}$ is to be fulfilled. In each round, the validators send their decision signed with their private key from their public-private ephemeral key set. Each node destroys its private key just after one time use.

The propagated messages $m^{r,1}_{i}$ and  $\{Head(B^{r}_{i}), \sigma^{r,1}_{i}\}$ for $n_{i} \in V_{r,1}$ initiate the beginning of step $(r,2)$. As per the followed convention, the set of validators for the round $(r,2)$ are given by $V_{r,2}$. In $(r,2)$, the validator $n_{j} \in V_{r,2}$ waits for a predetermined time period, then checks the different hash values received from the users in $V_{r,1}$. After the predetermined time period (network parameter), the validator $n_{j}$ checks all the credential values received by the selected nodes in $(r,1)$ and finds the $\sigma^{r,1}_{i}$, such that the normalized decimal value of its hash value, $\Theta^{r,1}_{i} < \Theta^{r,1}_{x}$ where $\forall i,x \in V_{r,1} \subset N$, and $x \neq i$. After determining the suitable proposer credential (say $\sigma^{r,1}_{i}$), if the data associated with $n_{i}$, i.e., components of $m^{r,1}_{i}$ are valid, the validator $n_{j} \in V_{r,2}$ sets its voting message as $vm^{r,2}_{j} = vm^{'} = \{H(B^{r}_{i}),i\}$ and propagates the message $m^{r,2}_{j}$ in the network for the validation in step $(r,3)$. The components of $m^{r,2}_{j}$ are $\{ SIG_{j_{j}}(vm^{r,2}_{j}), \sigma^{r,2}_{j}\}$. If the data associated with $\sigma^{r,1}_{i}$ is not valid, $n_{j} \in V_{r,2}$ propagates the vote for the empty block.

The subsequent steps have two components, graded consensus followed by binary consensus. In the step $(r,3)$, a new set of users $n_{k} \in V_{r,3}$ accumulate the messages from $V_{r,2}$ and evaluate the messages  $m^{r,2}_{j}$ received from $n_{j} \in V_{r,2}$  value to decide on which block to include. In step $(r,3)$, the validator $n_{k}$ with $k\in V_{r,3}$ (set of validators in $(r,3)$) propagates the message $m^{r,3}_{x} = \{ SIG_{k}(vm^{r,3}_{k}), \sigma^{r,3}_{k}\}$ into the network. The structure of $vm^{r,3}_{k}$ is same as that of $vm^{r,2}_{j}$. The choice for $vm^{r,3}_{x}$ is between a hash of non-empty block proposed in $(r,1)$ validated by atleast $t_{H}$ messages from $(r,2)$ or the hash of an empty block as default message after a predefined time period. 

The step $(r,4)$ and the subsequent rounds are binary consensus procedure, in which the decision is taken to select either the valid block or an empty block. In $(r,4)$, the validator $n_{a}$ with $a \in V_{r,4}$ (set of validators in $(r,4)$) accumulates the messages from $V_{r,3}$. The decision making is made binary by introducing two additional variables, $g^{r,4}_{a}$ and $b^{r,4}_{a}$ for $n_{a} \in V_{r,4}$. After evaluating all the valid messages among $m^{r,3}_{j}$ for all $n_{k} \in V_{r,3}$, $n_{a} \in V_{r,4}$ in has four choices. If there are at least $t_{H}$ valid messages advocating a non-empty block, then $g^{r,4}_{a} = 2$ and $b^{r,4}_{a} = 0$. For at least $t_{H}$ valid messages advocating an empty block, then $g^{r,4}_{a} = 0$ and $b^{r,4}_{a} = 1$. Otherwise, if there are at least $\left \lceil{\frac{t_{H}}{2}}\right \rceil$ valid messages advocating some other non-empty block, the values are set as $g^{r,4}_{a} = 1$ and $b^{r,4}_{a} = 1$. For all the other cases, $n_{a}$ sets $g^{r,4}_{a} = 0$ and $b^{r,4}_{a} = 1$. The message propagated by $n_{a} \in V_{r,4}$ after $(r,4)$ is of the form $m^{r,4}_{a} = \{ SIG_{a}(b^{r,4}_{a}), SIG_{a}(vm^{r,4}_{a}), \sigma^{r,4}_{a}\}$. The resultant block is finalised in subsequent steps based on the number of $b^{r,s}_{a}$ values with $1$ value means advocating empty block and $0$ value means voting for non-empty block. Ideally, in step $(r,5)$ the block $r$ is finalised. However, if that is not the case there are further steps defined to carry out consensus process  through coin-flip based protocol till the network has at least $t_{H}$ confirmed votes of either of the values $1$ or $0$.

\end{document}